\documentclass[twocolumn,showpacs,preprintnumbers,amsmath,amssymb]{revtex4}

\usepackage{graphicx}

\begin{document}

\title{Parallels between the dynamics at the noise-perturbed onset of chaos
in logistic maps and the dynamics of glass formation}
\author{F. Baldovin$^{1,2,3}$\thanks{email: baldovin@pd.infn.it} 
and A. Robledo$^{1}$\thanks{email: robledo@fisica.unam.mx},}
\affiliation{$^{1}$Instituto de F\'{\i}sica,\\
Universidad Nacional Aut\'{o}noma de M\'{e}xico,\\
Apartado Postal 20-364, M\'{e}xico 01000 D.F., Mexico\\
$^2$ INFM-Dipartimento di Fisica, Universit\`a di Padova,\\
Via Marzolo 8, I-35131 Padova, Italy\\
$^3$ Sezione INFN, Universit\`a di Padova, \\
Via Marzolo 8, I-35131 Padova, Italy }

\date{\today }

\begin{abstract}
We develop the characterization of the dynamics at the noise-perturbed edge
of chaos in logistic maps in terms of the quantities normally used to
describe glassy properties in structural glass formers. Following the
recognition [Phys. Lett. \textbf{A 328}, 467 (2004)] that the dynamics at
this critical attractor exhibits analogies with that observed in thermal
systems close to vitrification, we determine the modifications that take
place with decreasing noise amplitude in ensemble and time averaged
correlations and in diffusivity. We corroborate explicitly the occurrence of
two-step relaxation, aging with its characteristic scaling property, and
subdiffusion and arrest for this system. We also discuss features that
appear to be specific of the map.

\end{abstract}

\maketitle

\section{ Introduction}

The erratic motion of a Brownian particle is usually described by the
Langevin theory \cite{chaikin&lubensky1}. As it is well known, this method
finds a way round the detailed consideration of many degrees of freedom by
representing via a noise source the effect of collisions with molecules in
the fluid in which the particle moves. The approach to thermal equilibrium
is produced by random forces, and these are sufficient to determine
dynamical correlations, diffusion, and a basic form for the
fluctuation-dissipation theorem \cite{chaikin&lubensky1}.

In the same spirit, attractors of nonlinear low-dimensional maps under the
effect of external noise can be used to model states in systems with many
degrees of freedom. In a one-dimensional map with only one control parameter 
$\mu $ the consideration of external noise could be thought to represent the
effect of many other systems coupled to it, like in the so-called coupled
map lattices \cite{kaneko1}. Notice that the general map formula 
\begin{equation}
x_{t+1}=x_{t}+h_{\mu }(x_{t})+\sigma \xi _{t},  \label{mapnoise1}
\end{equation}
is a discrete form for a Langevin equation with nonlinear `friction force'
term $h_{\mu }$. In Eq. (\ref{mapnoise1}) $t=0,1,\ldots $ is the iteration
time, $\xi _{t}$ is a Gaussian white noise ($\langle \xi _{t}\xi _{t^{\prime
}}\rangle =\delta _{t,t^{\prime }}$), and $\sigma $ measures the noise
intensity.

An interesting option offered by Eq. (\ref{mapnoise1}) is the study of
singular states known to exhibit anomalous dynamics. For instance, the
so-called onset of chaos in logistic maps is a critical attractor with
nonergodic and nonmixing phase-space properties, and the perturbation of
this attractor with noise transforms its trajectories into genuinely chaotic
ones with regular ergodic and mixing properties. A case in point here would
be to obtain the atypical dynamics near an ergodic to nonergodic transition
and compare it, for example, to that in supercooled liquids close to glass
formation, where also an ergodic to nonergodic transition is believed to
occur. The dynamical properties for such glassy states appear associated to
two-time correlations with loss of time translation invariance (TTI) and
aging scaling properties, as well as to subdiffusion and arrest 
\cite{debenedetti1,angell1}. 
The specific question we would like to address is
whether there are properties shared, and if so, to what extent, by critical
attractors in nonlinear low-dimensional maps and nonergodic states in
systems with many degrees of freedom.

\begin{figure}
\begin{center}
\includegraphics[width=0.95\columnwidth]{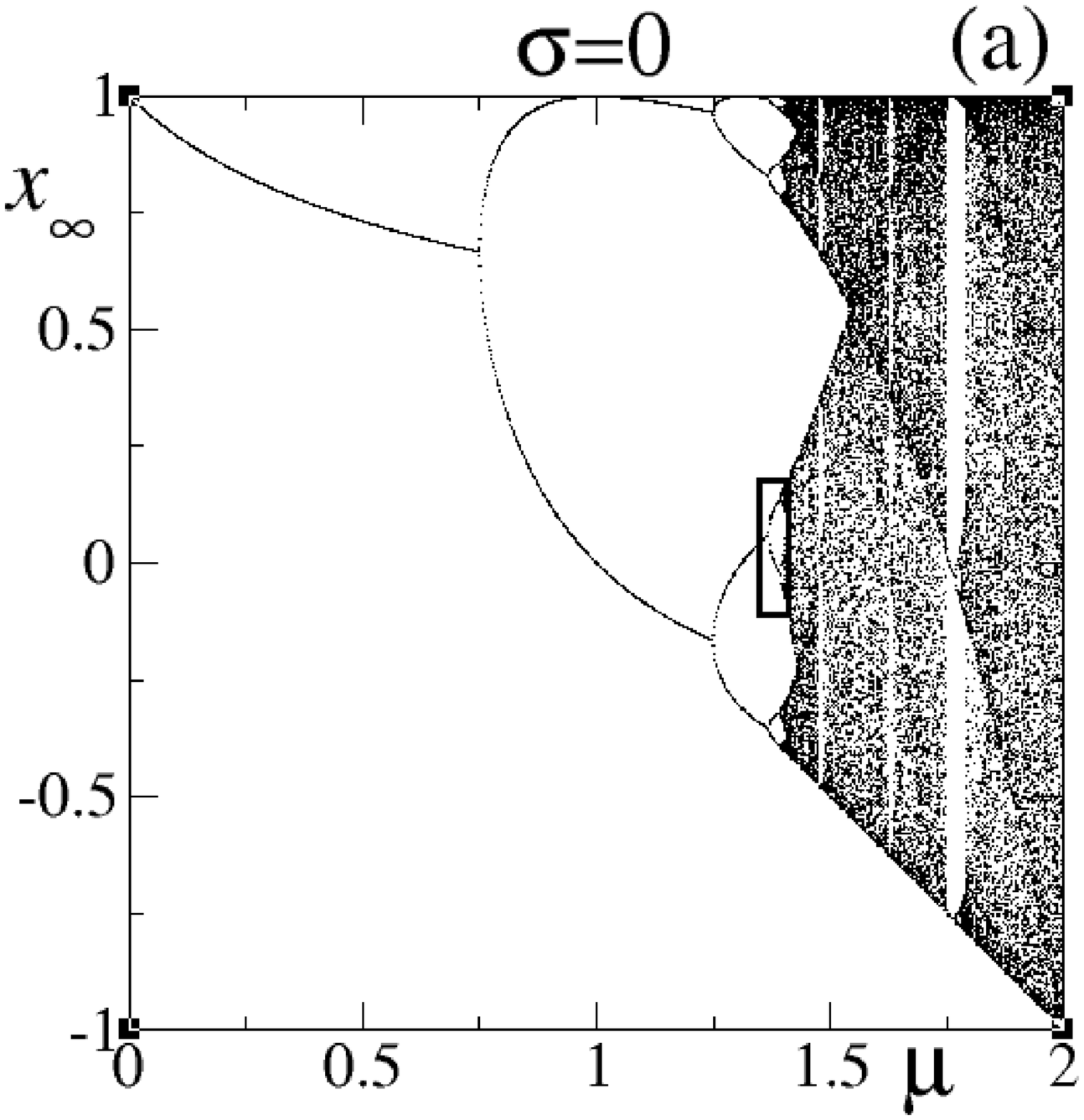} 
\includegraphics[width=0.95\columnwidth]{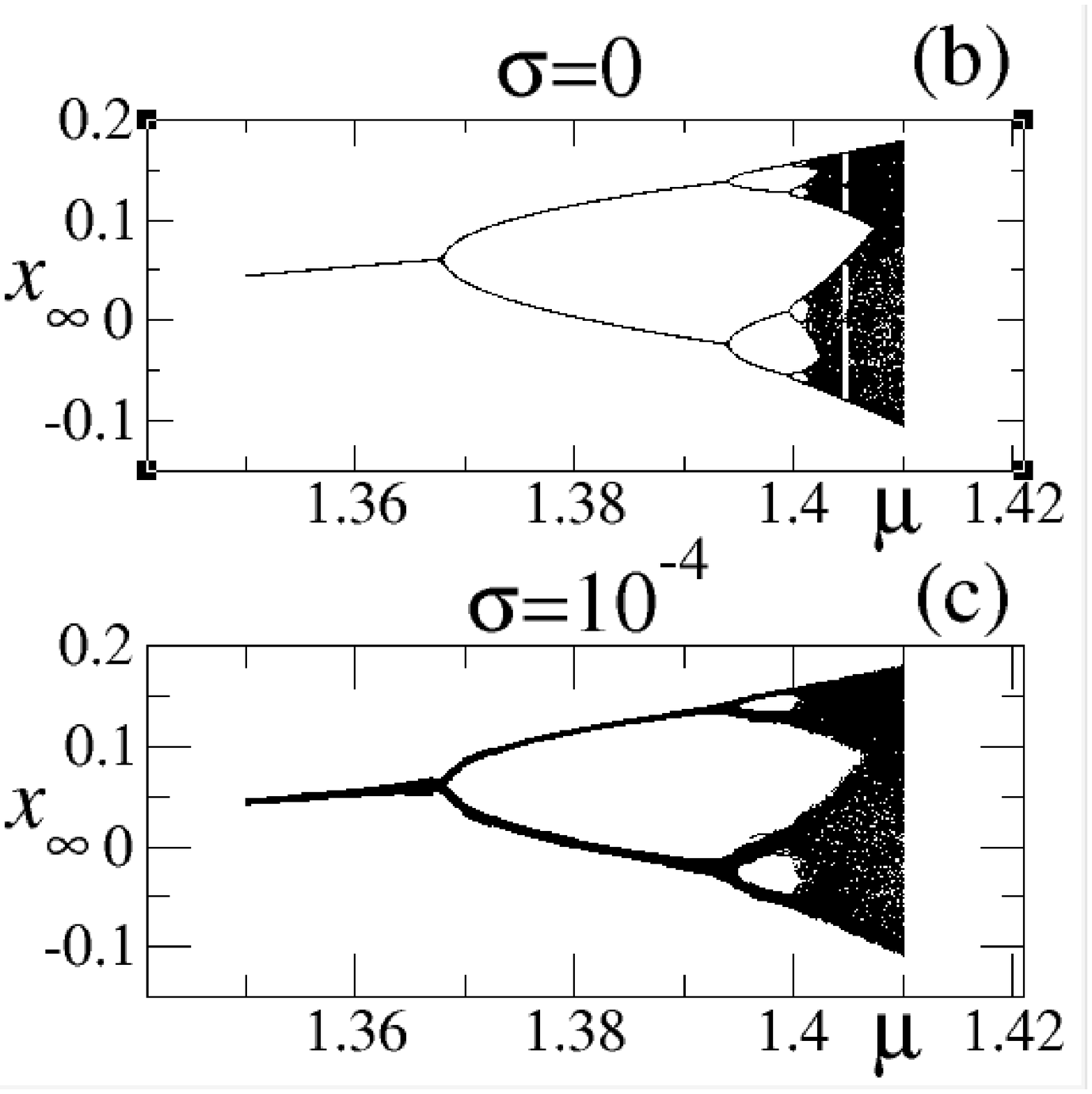}
\end{center}
\caption{ (a) Logistic map attractor. (b) Magnification of the box in (a).
(c) Noise-induced bifurcation gap in the magnified box. }
\label{fig_att_lpn}
\end{figure}

Recently \cite{robglass1}, it has been realized that the dynamics at the
noise-perturbed edge of chaos in logistic maps shows similarities with that
observed in supercooled liquids close to vitrification. Three major features
of glassy dynamics in structural glass formers, two-step relaxation, aging,
and a relationship between relaxation time and configurational entropy, were
shown to be displayed by the properties of orbits with vanishing Lyapunov
coefficient. Interestingly, the previously known properties in
control-parameter space of the noise-induced bifurcation gap 
(See Fig. \ref{fig_att_lpn}) \cite{schuster1,crutchfield1} 
play a central role in
determining the characteristics of dynamical relaxation at the chaos
threshold, and this was exploited to uncover the analogy between the
dynamical and the thermal systems in which the noise amplitude $\sigma $
plays a role equivalent to a temperature difference $T-T_{g}$, where $T_{g}$
is the so-called glass transition temperature \cite{debenedetti1,angell1}.

In Ref. \cite{robglass1} only the properties of single-time functions (i.e.
trajectories) were discussed and here we focus the analysis on two-time
correlations. This would allow for a closer examination of the analogy with
glassy dynamics as the study of the latter often centers on two-time
correlations. However, we also look at diffusion properties via the
determination of the mean square displacement of trajectories in a suitably
space-extended map. To lay emphasis on the similarities in the dynamics
between the two types of systems we analyze the attainment of TTI in the
correlations caused by the action of noise. We recall 
\cite{beale1,grassberger1} 
that exposure of attractors to noise have the features
of an activated process a mechanism that is usually considered in the
interpretation of relaxation processes in glass formers. Also we make up a
simple 'landscape' picture for the properties of the noise-perturbed map in
order to compare it with that obtained from the multidimensional energy
landscape of supercooled liquids. Aware of our crude attempt to contrast the
dynamics of a single map (although equivalent to a coupled array of such
maps when described via a Langevin-type equation) with that of thermal
systems we point out the main differences encountered. Specifically, the
dynamics at the onset of chaos displays regular patterns absent in the known
(experimental or computed) dynamics of molecular systems. These differences
reflect the peculiarities of the period-doubling route to chaos displayed by
unimodal maps.

The structure of the body of the article is as follows: In Section 2 we
recall essential properties of the logistic map under additive noise, e.g.
the bifurcation gap and its time-dependent manifestation at the chaos
threshold. In Section 3 we present results on ensemble-averaged two-time
correlations for $\sigma \gtrsim 0$, the attainment of TTI, and the
development of two-step relaxation as $\sigma \rightarrow 0$. In Section 4
we focus on time-averaged two-time correlations and the occurrence of the
characteristic aging scaling at the onset of chaos with $\sigma =0$. In
Section 5 we present a repeated-cell map for diffusion at the onset of chaos
and show the evolution in behavior from diffusive to subdiffusive and
finally to localization as $\sigma \rightarrow 0$. In Section 6 we make
final remarks.

\section{Noise-perturbed onset of chaos}
\begin{figure}
\begin{center}
\includegraphics[width=0.95\columnwidth]{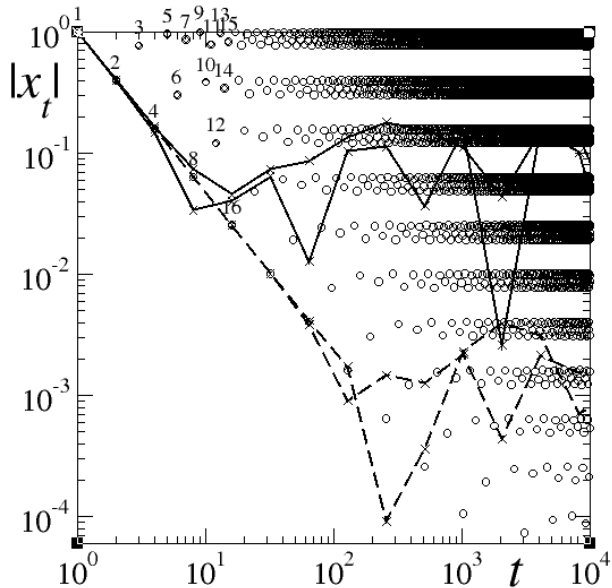}
\end{center}
\caption{Absolute values of positions in logarithmic scales of iterations $t$
for various trajectories at the onset of chaos 
$\protect\mu _{c}(\protect\sigma )$ starting at $x_{0}=0$. 
Empty circles correspond to 
$\protect\sigma=0$ where the numbers label time $t=1,.,16$. Full (and dashed) lines
represent trajectories for noise amplitude $\protect\sigma =10^{-3}$ (and 
$\protect\sigma =10^{-6}$) plotted only at times $t=2^{n}$. }
\label{fig_phsp_lpn}
\end{figure}

We describe now the effect of additive noise in the dynamics at the onset of
chaos in the logistic map 
\begin{equation}
x_{t+1}=1-\mu x_{t}^{2}+\sigma \xi _{t},\ -1\leq x_{t}\leq 1,\ 0\leq \mu
\leq 2,  \label{logisticnoise1}
\end{equation}
where $h_{\mu }$ in Eq. (\ref{mapnoise1}) has taken the form $h_{\mu
}(x)=1-x-\mu x^{2}$. For $\sigma =0$ at the onset of chaos at $\mu
_{c}=1.40115...$ the orbit with attractor initial condition $x_{0}=0$
consists of positions arranged as nested power-law time subsequences that
asymptotically reproduce the full period-doubling cascade that occurs for 
$\mu <\mu _{c}$ \cite{baldovin1,baldovin2}. See empty circles in 
Fig. \ref{fig_phsp_lpn}. 
This orbit is the last (the accumulation point) of the
so-called `superstable' orbits of period $2^{n}$\ which occur at 
$\mu =\overline{\mu }_{n}<$ $\mu _{c}$, $n=1,2,...$, a superstable orbit of period 
$2^{\infty }$. Superstable orbits include $x=0$ as one of their positions
and their Lyapunov exponent $\lambda _{1}$ diverges to $-\infty $ 
\cite{schuster1}. A transient dynamics at the onset of chaos is observed for
trajectories with initial position outside the Feigenbaum attractor but we
shall not consider this in what follows. For $\sigma >0$ the noise
fluctuations wipe the fine structure of the periodic attractors as the
iterate visits positions within a set of bands or segments similar to those
in the chaotic attractors, nevertheless there remains a well-defined
transition to chaos at $\mu _{c}(\sigma )$ where the Lyapunov exponent 
$\lambda _{1}$ changes sign \cite{schuster1,crutchfield1}. The period
doubling of bands ends at a finite maximum period $2^{N(\sigma )}$ as $\mu
\rightarrow \mu _{c}(\sigma )$ (see Fig. \ref{fig_att_lpn}c) and then
decreases at the other side of the transition. This effect displays scaling
features and is referred to as the bifurcation gap 
\cite{schuster1,crutchfield1}. For instance, $\Delta \mu _{c}\sim \sigma ^{\gamma
}$ where $\Delta \mu \equiv \mu _{c}(0)-\mu _{c}(\sigma )$ and $\gamma =\ln
\delta _{F}/\ln \nu $, where $\delta _{F}=0.46692...$ is one of the two
Feigenbaum's universal constants (the 2nd, $\alpha _{F}=2.50290...$,
measures the power-law period-doubling spreading of iterate positions), and 
$\nu \simeq 2\sqrt{2}\alpha _{F}(1+1/\alpha _{F}^{2})^{-1/2}\simeq 6.619$.
See \cite{robglass1} and references therein. When $\sigma $ is small the
trajectories visit sequentially the set of $2^{N(\sigma )}$ disjoint bands
leading to a cycle, but the behavior inside each band is irregular. These
trajectories represent ergodic states as the accessible positions have a
fractal dimension equal to the dimension of phase space. When $\sigma =0$
the trajectories correspond to a nonergodic state, since as $t\rightarrow
\infty $ the positions form only a Cantor set of fractal dimension 
$d_{f}=0.538...$. Thus the removal of the noise $\sigma \rightarrow 0$ leads
to an ergodic to nonergodic transition in the map.

As shown in Ref. \cite{robglass1} when $\mu =\mu _{c}(\sigma )$ 
($\sigma >0$) there is a `crossover' or `relaxation' time $t_{x}=\sigma ^{r-1}$, 
$r=1-\ln 2/\ln \nu \simeq 0.6332$, between two different time evolution
regimes. This crossover occurs when the noise fluctuations begin suppressing
the fine structure of the attractor as displayed by the superstable orbit
with $x_{0}=0$ described above. (See full and dashed lines in 
Fig. \ref{fig_phsp_lpn}). 
For $t<t_{x}$ the fluctuations are smaller than the
distances between the neighboring subsequence positions of the $x_{0}=0$
orbit at $\mu _{c}(0)$, and the iterate position with $\sigma >0$ falls
within a small band around the $\sigma =0$ position for that $t$. The bands
for successive times do not overlap. Time evolution follows a subsequence
pattern close to that in the noiseless case. When $t\sim t_{x}$ the width of
the noise-generated band reached at time $t_{x}=2^{N(\sigma )}$ matches the
distance between adjacent positions, and this implies a cutoff in the
progress along the position subsequences. At longer times $t>t_{x}$ the
orbits no longer trace the precise period-doubling structure of the
attractor. The iterates now follow increasingly chaotic trajectories as
bands merge with time. This is the dynamical image -- observed along the
time evolution for the orbits of a single state $\mu _{c}(\sigma )$ -- of
the static bifurcation gap initially described in terms of the variation of
the control parameter $\mu $ \cite{crutchfield1}.

The entropy associated to the distribution of the iterate positions within
the $2^{N}$ bands has the form $S_{c}=2^{N}\sigma s$, where $s$ is the
entropy associated to a single band. Use of $2^{N}=t_{x}$ and $t_{x}=\sigma
^{r-1}$, $r-1\simeq -0.3668$ \cite{robglass1}, leads to 
\begin{equation}
t_{x}=(s/S_{c})^{(1-r)/r}.
\end{equation}
Since $(1-r)/r\simeq 0.5792$ then $t_{x}\rightarrow \infty $ and 
$S_{c}\rightarrow 0$ as $\sigma \rightarrow 0$. See \cite{robglass1} for
details on the derivation. We have compared \cite{robglass1} this expression
with its counterpart in structural glass formers, the Adam-Gibbs equation 
\cite{debenedetti1}.

\begin{figure}
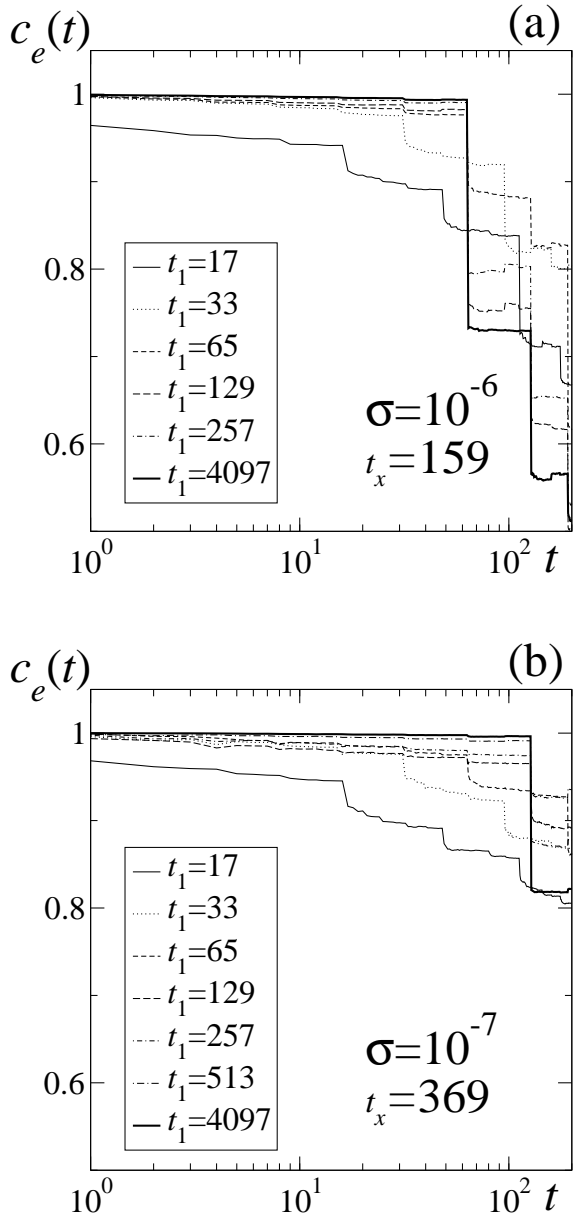

\begin{center}
\includegraphics[width=0.95\columnwidth]{alpha_therm_1em6.eps}\\
\vspace{0.25cm}
\includegraphics[width=0.95\columnwidth]{alpha_therm_1em7.eps}
\end{center}
\caption{Development of time translation invariance (TTI) in the correlation 
$c_{e}(t_{1},t=t_{2}-t_{1})$ through the action of noise of amplitude 
$\protect\sigma $ at the onset of chaos. All trajectories start $x_{0}=0$ and 
$t_{x}$ is the time to reach the bifurcation gap. }
\label{fig_alpha_therm}
\end{figure}

\section{ Two-step relaxation}
The time evolution of equilibrium two-time correlations in supercooled
liquids on approach to glass formation display a two-step process of
relaxation. This consists of a primary power-law decay in time difference 
$t=t_{2}-t_{1}$ that leads into a plateau, and at the end of this there is a
second power law decay that evolves into a faster decay that can be fitted
by a stretched exponential \cite{debenedetti1}. Also, the duration $t_{x}$
of the plateau increases as an inverse power law of the difference 
$T-T_{g}\gtrsim 0$ as the temperature $T$ decreases to the glass transition
temperature $T_{g}$. The first and second decays are usually referred to as
the $\beta $ and the $\alpha $ relaxation processes, respectively 
\cite{debenedetti1}. An observable example of such correlation function, both
experimentally and numerically, is the Fourier transform of the
density-density correlation at time difference $t$. The former is known as
the intermediate scattering function while the latter is known as the van
Hove function \cite{debenedetti1}.

\begin{figure}
\begin{center}
\includegraphics[width=0.95\columnwidth]{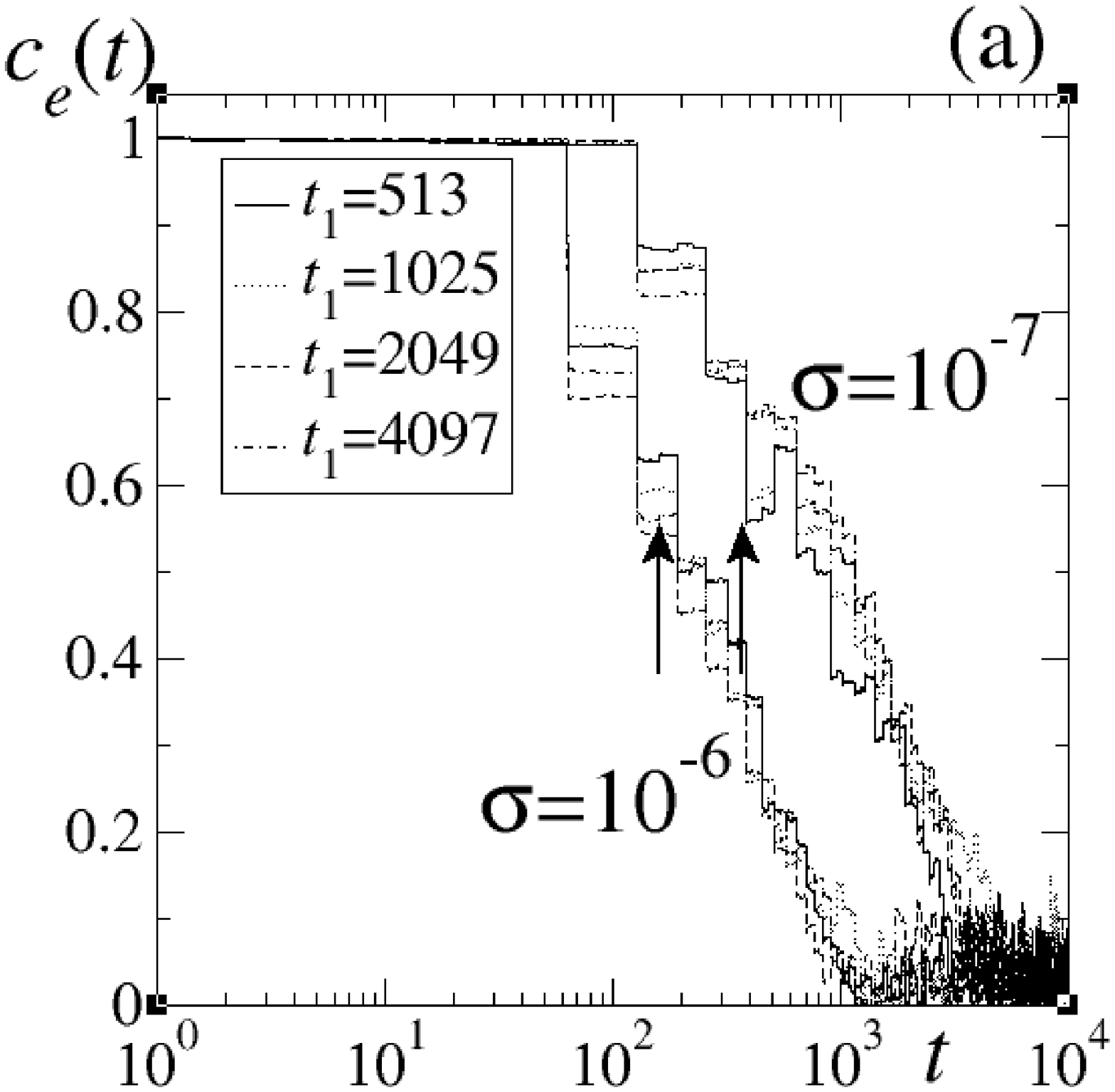}\\
\vspace{0.25cm}
\includegraphics[width=0.95\columnwidth]{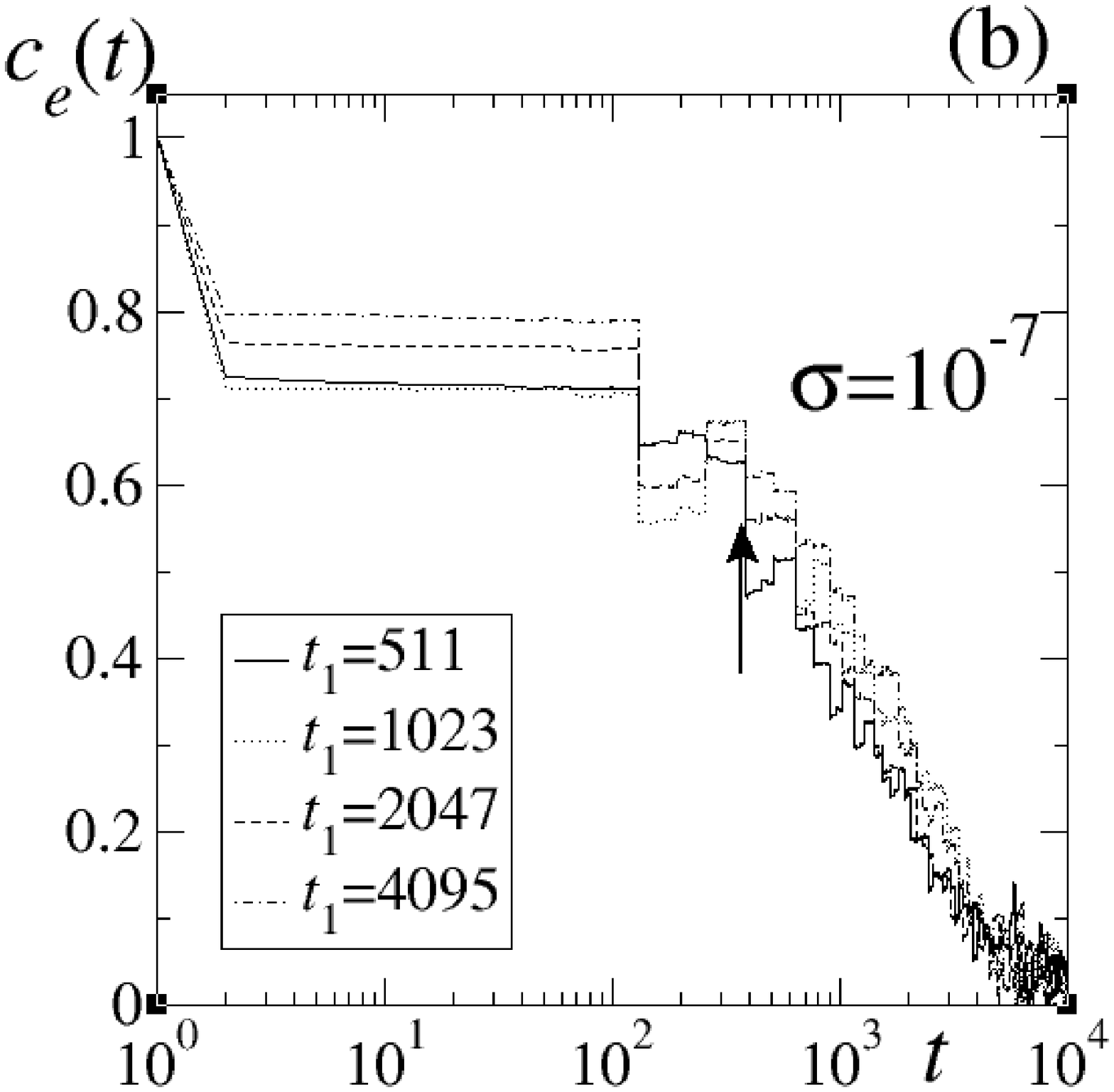} 
\end{center}
\caption{ Relaxation at the onset of chaos according to the correlation
function $c_{e}(t_{1},t=t_{2}-t_{1})$ defined in Eq. 
(\protect\ref{correlation1}) for an ensemble of $500$ trajectories starting at $x_{0}=0$. 
In (a) $t_{1}$ is of the form $2^{n}+1$
and the noise amplitude is $\protect\sigma =10^{-6}$
and $10^{-7}$. 
Vertical arrows, from left to right, indicate the crossover time 
$t_{x}(\protect\sigma )$. 
In (b) $t_{1}=2^{n}-1$ and $\protect\sigma =10^{-7}$.
}
\label{fig_beta_alpha}
\end{figure}

The study of single trajectory properties (one time functions) in Ref. 
\cite{robglass1} led to the suggestion that the dynamical behavior in the map at 
$\mu _{c}(\sigma )$ would show parallels to the relaxation properties of
glass formers. For instance, the analog of the $\beta $ relaxation would be
obtained by considering initial conditions $x_{0}$ outside the critical
attractor since the ensuing orbits display a power-law transient as the
positions approach asymptotically those of the attractor. The intermediate
plateau would correspond to the regime $t<t_{x}$, described in the previous
section, when the iterates are confined to nonintersecting bands before they
reach the bifurcation gap, its duration $t_{x}$ grows as an inverse power
law of $\sigma $. The analog of the $\alpha $ relaxation was proposed to be
the band merging crossover process that takes place for $t>t_{x}$. To
explore more closely these similarities we evaluated the two-time
correlation function 
\begin{equation}
c_{e}(t_{1},t_{2})=\frac{\left\langle x_{t_{2}}x_{t_{1}}\right\rangle
-\left\langle x_{t_{2}}\right\rangle \left\langle x_{t_{1}}\right\rangle }
{\chi _{t_{1}}\chi _{t_{2}}},\;1\leq t_{1}\leq t_{2}  \label{correlation1}
\end{equation}
for different values of the noise amplitude $\sigma $. 
In Eq. (\ref{correlation1}) 
$\left\langle ...\right\rangle $ represents an average over
an ensemble of trajectories all of them starting with initial conditions 
$x_{0}=0$ and $\chi _{t_{i}}=\sqrt{\left\langle x_{t_{i}}^{2}\right\rangle
-\left\langle x_{t_{i}}\right\rangle ^{2}}$.

We first address the question of whether the exposure of trajectories to
noise has the effect of introducing, after an initial transient period, a
time translation invariance (TTI) property into the correlation 
in Eq. (\ref{correlation1}), i.e. 
$c_{e}(t_{1},t_{2})\simeq c_{e}(t=t_{2}-t_{1})$, 
$t_{1} $ large. The presence of this effect would be analogous to
thermalization in molecular systems after which equilibrium correlations are
measured or computed (in glass forming systems for $T>T_{g}$). 
In Fig. \ref{fig_alpha_therm} we
show how TTI develops and is maintained for a sufficiently large time
difference interval $t=t_{2}-t_{1}$. The numerical limitation in evaluating
accurate values for $\mu _{c}(\sigma )$ leads to an upper bound for $t$, but
we checked that increasing precision in $\mu _{c}(\sigma )$ leads to a
larger interval for $t $ for which TTI is observed.\ In view of the results
shown in Fig. \ref{fig_alpha_therm} 
we can conclude that under external noise of weak amplitude
the ensemble of trajectories 'thermalizes' asymptotically into an
'equilibrium' attractor.

The TTI property in the map still retains certain memory of the initial 
$t_{1}$, only in a generic way that reflects the characteristic symmetries of
the period-doubling onset of chaos of the logistic map. Notice that in Fig 3
the values for $t_{1}$ are of the form $t_{1}=2^{n}+1$. It is found that
both the initial decay and the value of the intermediate plateau of 
$c_{e}(t) $ are fixed when the values of $t_{1}$ belong to the sequence 
$t_{1}=2^{n}+k$, $n$ large with $k$ fixed. However the main decay process of 
$c_{e}(t)$ appears to be independent of $t_{1}$. In Figs. \ref{fig_beta_alpha}
a and \ref{fig_beta_alpha}b we show the behavior of $c_{e}(t)$,
respectively, when $t_{1}=2^{n}+1$ and $t_{1}=2^{n}-1$, $n=9,10,...$ In the
first case (Fig. \ref{fig_beta_alpha}a) the correlation maintains the
initial value of unity from $t=0$ until the main decay process sets in,
while in the second case (Fig. \ref{fig_beta_alpha}b) there is initial decay
at short times followed by a plateau that ends when the same main decay as
in Fig. \ref{fig_beta_alpha}a takes place. For other values of $k$ the
correlation $c_{e}(t)$ shows a behavior similar to that in 
Fig. (\ref{fig_beta_alpha}b) with different values for the duration of the initial
decay (that we refer to as the $\beta $ decay) and for the plateau, but
always with the same main decay (that we refer to as the $\alpha $ decay).
The $\alpha $ decay is itself made up of several plateaus the values of
which alternate when $n$ is varied (as shown in Figs. \ref{fig_beta_alpha}a
and \ref{fig_beta_alpha}b). In all cases the duration of the main plateau
coincides approximatively with the crossover time $t_{x}$ at which the
bifurcation gap is reached. (See the vertical arrows in 
Figs. \ref{fig_beta_alpha}a and \ref{fig_beta_alpha}b). Thus, the identification of
the encounter of the bifurcation gap as the triggering event of the $\alpha $
relaxation process \cite{robglass1} seems to be confirmed by the numerical
evaluation of $c_{e}(t)$.

\section{Aging}

In glass forming systems when $T-T_{g}\lesssim 0$ (nonequilibrium) two-time
correlations lose time translation invariance, and the dependence on the two
times $t_{1}$ and $t_{2}=t+t_{1}$, has the characteristic known as aging 
\cite{angell1}. More specifically, the time scale for the response to an
external perturbation increases with the waiting time $t_{w}$, the time
interval $t_{w}=$ $t_{1}-t_{0}$ from system preparation at $t_{0}=0$ to the
moment of the perturbation at $t_{1}$. As a consequence, the equilibrium
fluctuation-dissipation relation that relates response and correlation
functions breaks down \cite{angell1}. In this regime the decay of response
and correlation functions display a scaling dependence on the ratio $t/t_{w}$
\cite{angell1}.

\begin{figure}
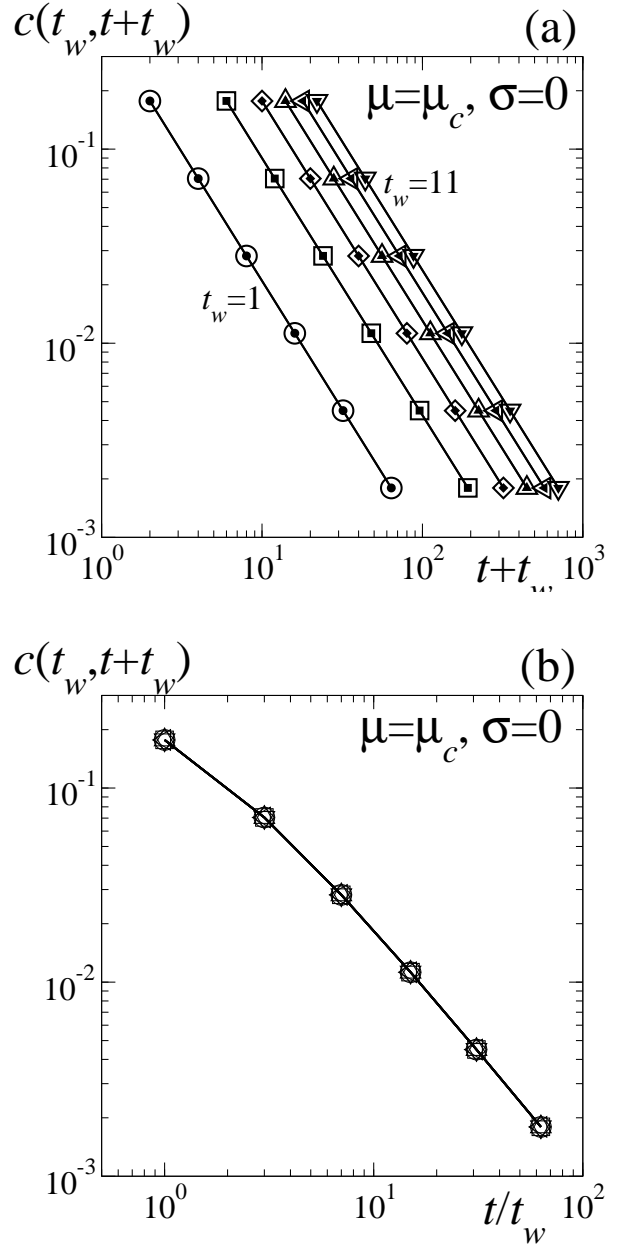

\begin{center}
\includegraphics[width=0.95\columnwidth]{corr_time_a.eps}\\
\vspace{0.25cm}
\includegraphics[width=0.95\columnwidth]{corr_time_b.eps}
\end{center}
\caption{ Aging according to the correlation $c(t_{w},t+t_{w})$ given by Eq.
(\protect\ref{correlation2}) for the Feigenbaum attractor 
($\protect\mu=\protect\mu_c,\;\protect\sigma=0$). Total observation time is $n=1000$. In
(a) is shown the explicit dependence on the waiting time (from left to right 
$t_w=1,\;3,\;5,\;7,\;9,\;11$). In (b) all curves collapse upon rescaling 
$t/t_w$. }
\label{fig_corr_time}
\end{figure}

As indicated in Ref. \cite{robglass1} the power-law position subsequences
shown in Fig. \ref{fig_phsp_lpn} that constitute the superstable orbit of
period $2^{\infty }$ within the noiseless attractor at $\mu _{c}(0)$ imply a
built-in aging scaling property for the single-time function $x_{t}$. These
subsequences are relevant for the description of trajectories that are at
first held at a given attractor position for a waiting period of time $t_{w}$
and then released to the normal iterative procedure. We chose the holding
positions to be any of those along the top band shown in 
Fig. \ref{fig_phsp_lpn} with $t_{w}=2k+1$, $k=0,1,...$. One obtains \cite{robglass1} 
\begin{equation}
x_{t+t_{w}}\simeq \exp _{q}(-\lambda _{q}t/t_{w}),  \label{trajectory1}
\end{equation}
where $\exp _{q}(x)\equiv \lbrack 1-(q-1)x]^{1/1-q}$ and $\lambda _{q}=\ln
\alpha _{F}/\ln 2$. This property is gradually removed when noise is turned
on. The presence of a bifurcation gap limits its range of validity to total
times $t_{w}+t$ $<t_{x}(\sigma )$ and so progressively disappears as $\sigma 
$ is increased \cite{robglass1}.

When $\sigma =0$ trajectories are nonergodic and ensemble and time averages
are not equivalent. For this reason we use a time-averaged correlation 
$c(t_{1},t_{2})$ to study aging and its related scaling property at the onset
of chaos for $\sigma =0$, instead of the ensemble-averaged 
$c_{e}(t_{1},t_{2})$ in Eq. (\ref{correlation1}). Also for this case 
$c_{e}(t_{1},t_{2})$ is not defined as $\chi _{t_{i}}=0$ when the initial
positions are all $x_{0}=0$ or $x_{t_{w}}$. We chose for $c(t_{1},t_{2})$
the form 
\begin{equation}
c(t_{w},t+t_{w})=(1/N)\sum_{n=n_{0}}^{N}\phi ^{(n)}(t_{w})\phi
^{(n)}(t+t_{w}),  \label{correlation2}
\end{equation}
where $\phi (t)=f_{\mu _{c}}^{(t)}(0)$ and $f_{\mu }(x)=1-\mu x^{2}$, and
with $n_{0}$ any positive integer and $N\gg n_{0}$ a large integer. This
definition of $c(t_{w},t+t_{w})$ is designed to capture the power-law
patterns of the trajectories at the noiseless onset of chaos. 
Eq. (\ref{correlation2}) considers multiples of the two reference times $t_{w}$ and 
$t+t_{w}$, i.e. times at which trajectories recurrently visit a given region
of the attractor \cite{robmori1}. In Fig. \ref{fig_corr_time}a we show 
$c(t_{w},t+t_{w})$ for different values of $t_{w}$, and in 
Fig. \ref{fig_corr_time}b the same data where the rescaled variable 
$t/t_{w}=2^{n}-1$, $t_{w}=2k+1$, $k=0,1,...$, has been used. We have calculated 
$c(t_{w},t+t_{w})$ for different values of $N$ and $n_{0}$ and found in both
cases the same result as in Fig. \ref{fig_corr_time}a. The characteristic
scaling of aging behavior is especially clear.

\section{Subdiffusion and arrest.}

The sharp slow down of dynamics in supercooled liquids on approach to
vitrification is illustrated by the progression from normal diffusiveness to
subdiffusive behavior and finally to a halt in the growth of the molecular
mean square displacement within a small range of temperatures or densities 
\cite{weeks1,dawson1}. This deceleration of the dynamics is caused by the
confinement of any given molecule by a `cage' formed by its neighbors; and
it is the breakup and rearrangement of the cages which drives structural
relaxation, letting molecules diffuse throughout the system. Evidence
indicates that lifetime of the cages increases as conditions move toward the
glass transition, probably because cage rearrangements involve a larger
number of molecules as the glass transition is approached 
\cite{weeks1,dawson1}.

\begin{figure}
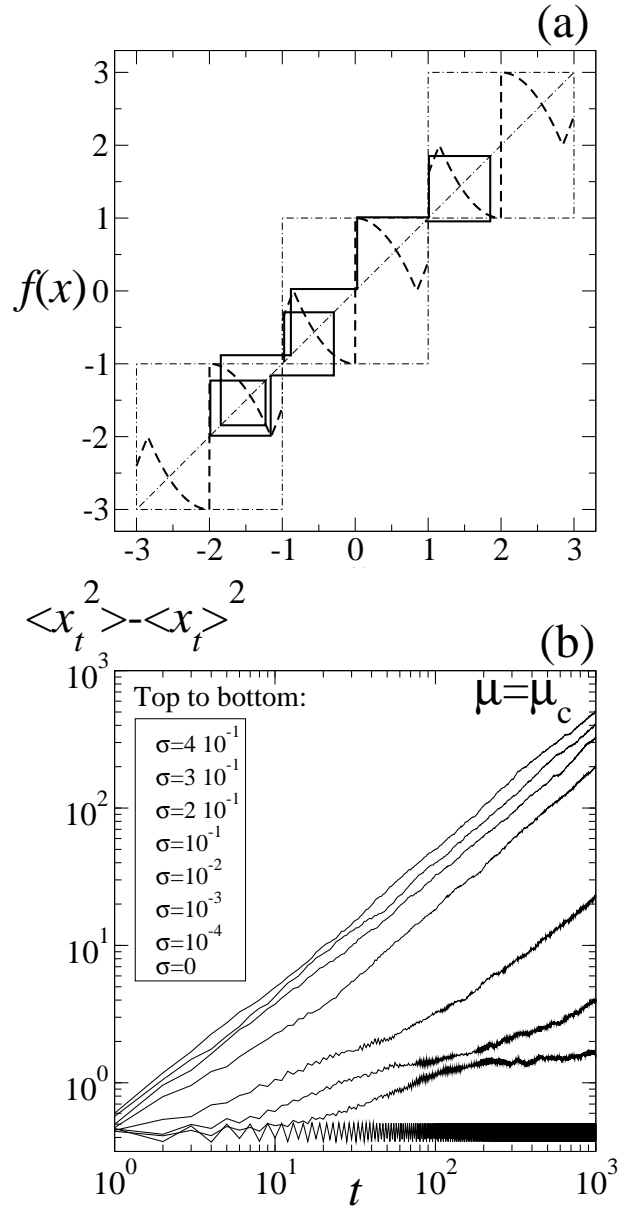

\begin{center}
\includegraphics[width=0.95\columnwidth]{diff_cell.eps}
\includegraphics[width=0.95\columnwidth]{diff_ensemble.eps}
\end{center}
\caption{ Glassy diffusion in the noise-perturbed logistic map. (a)
Repeated-cell map (thick dashed line) and trajectory (full line). (b) Time
evolution of the mean square displacement $\left\langle
x_{t}^{2}\right\rangle-\left\langle x_{t}\right\rangle ^{2}$ for an ensemble
of $1000$ trajectories with initial conditions randomly distributed inside 
$[-1,1]$. Curves are labeled by the value of the noise amplitude. }
\label{fig_diff}
\end{figure}

To investigate this aspect of vitrification in the map at 
$\mu _{c}(\sigma )$, 
we constructed a periodic map via repetition of a single (cell) map. This
setting has being used to study deterministic diffusion in nonlinear maps,
in which the trajectories migrate into neighboring cells due to chaotic
motion. For fully chaotic maps diffusion is normal \cite{schuster1} but for
marginally chaotic maps it is anomalous \cite{barkai1}. In our case we
design the map in such a way that diffusion is due only to the random noise
term, otherwise motion is confined to a single cell. So, we have the
periodic map $x_{t+1}=F(x_{t})$, $F(l+x)=l+F(x)$, $l=...-1,0,1,...$, where 
\begin{equation}
F(x)=\left\{ 
\begin{array}{c}
-\left\vert 1-\mu _{c}x^{2}\right\vert +\sigma \xi ,\;-1\leq x<0, \\ 
\left\vert 1-\mu _{c}x^{2}\right\vert +\sigma \xi ,\;0\leq x<1.
\end{array}
\right.  \label{cellmap1}
\end{equation}

Fig. \ref{fig_diff}a shows the repeated-cell map together with a portion of
one of its trajectories. As it can be observed, the escape from the central
cell into any of its neighbors occurs when $\left\vert F(x)\right\vert >1$
and this can only happen when $\sigma >0$. As $\sigma \rightarrow 0$ the
escape positions are confined to values of $x$ increasingly closer to $x=0$,
and for $\sigma =0$ the iterate position is trapped within the cell.
Likewise for any other cell. Fig. \ref{fig_diff}b shows the mean square
displacement $\left\langle x_{t}^{2}\right\rangle -\left\langle
x_{t}\right\rangle ^{2}$ as obtained from an ensemble of trajectories
initially distributed within the interval $[-1,1]$ for several values of
noise amplitude. The progression from normal diffusion to subdiffusion and
to final arrest can be plainly observed as $\sigma \rightarrow 0$. For small 
$\sigma $ ($\leq 10^{-2}$) $\left\langle x_{t}^{2}\right\rangle
-\left\langle x_{t}\right\rangle ^{2}$ shows a down turn and later an upturn
similar to those observed in colloidal glass experiments \cite{weeks1} and
attributed to cage rearrangements. In the map this feature reflects cell
crossings.

\section{Final remarks}

As we have shown, the dynamics of logistic maps at the chaos threshold in
the presence of noise displays elements reminiscent of glassy dynamics as
observed in molecular glass formers. The limit of vanishing noise amplitude 
$\sigma \rightarrow 0$ (the counterpart of the limit $T-T_{g}\rightarrow 0$
in the supercooled liquid) leads to loss of ergodicity. This nonergodic
state with vanishing Lyapunov coefficient $\lambda _{1}=0$ corresponds to
the limiting state, $\sigma \rightarrow 0$, $t_{x}\rightarrow \infty $, of a
family of small $\sigma $ states with properties reminiscent of those in
glass formers. Some additional comments may be useful to appreciate both
similarities and differences between the two types of systems.

Activated dynamics is a standard component in understanding the relaxation
mechanisms of glass formation \cite{debenedetti1,angell1} and so it is
pertinent to note that there is a similar characteristic in the dynamics
that we studied here. It has long been known \cite{beale1,grassberger1} that
the addition of external noise to a dissipative dynamical system (here a
nonlinear one-dimensional map) causes its trajectories to escape from
attractors. For chaotic attractors the mean escape time $T$ has an
exponential Arrhenius form
\[
T\approx T_{0}\exp (E_{0}/R), 
\]
where $E_{0}$ and $R$ are the minimum escape 'energy' and noise
'temperature', respectively, and $T_{0}$ is the inverse of the attempt rate.
The difference from the usual molecular set up is that the escape is from
the noiseless attractor and not from a minimum in the potential energy. For
critical attractors, like the onset of chaos, the minimum escape energy
vanishes and the escape time is expected to follow a power law with strong
fluctuations \cite{beale1,robmori1}. We have seen that as $\sigma
\rightarrow 0$ an increasing number of distinct but small phase space bands
are involved in an increasingly slow decay of dynamic correlations. On the
other hand, in the usual picture of activated dynamics in glass formers the
increase of the relaxation time as $T-T_{g}\rightarrow 0$ is thought to be
associated to an increase in the landscape energy barriers and the
assumption of some form of cooperative behavior by means of which a large
number of particles are rearranged through a very slow process 
\cite{debenedetti1}.

The structure of the phase space regions that are sampled by trajectories as
a function of the noise amplitude\ resemble the manner in which a glass
forming system samples its energy landscape as a function of temperature 
\cite{debenedetti1}. The numbers and widths of the attractor bands at noise
amplitude $\sigma $ can be thought to correspond to the numbers and extents
of potential energy basins of the landscape at a given depth set by\ the
temperature $T$. The 'landscape' for the attractor has certainly a very
simple structure when compared to that of the multidimensional energy
landscape in a molecular system. At $\sigma =0$ there is an infinite set of
minima that consists of all the points $M\rightarrow \infty $ of the
attractor, and as $\sigma $ increases these minima merge into finite sets of 
$M=2^{N(\sigma )}$ bands with $N$ decreasing as $\sigma $ grows. The regular
merging by two in the numbers $M$ and the features in the dynamics that they
imprint are properties specific of logistic type maps. Other kinds of
critical multifractal attractors, such as those for the critical circle map 
\cite{schuster1}, would exhibit other properties characteristic of the route
to chaos involved. This availability of detail would not be generally
present in the measurements or numerical computation of the dynamics of a
glass forming molecular system.

It is of interest to note that at $\mu _{c}(\sigma )$ the trajectories and
its resultant sensitivity to initial conditions are expressed for $t<t_{x}$
via the $q$-exponentials of the Tsallis $q$-statistics \cite{robglass1}. For 
$\sigma =0$ this analytical forms are exact \cite{baldovin1,robmori1} and an
identity linking accordingly generalized Lyapunov coefficients and rates of
entropy production, holds rigorously \cite{baldovin2,robmori1}. There is
nonuniform convergence related to the limits $\sigma \rightarrow 0$ and 
$t\rightarrow \infty $. If $\sigma \rightarrow 0$ is taken before 
$t\rightarrow \infty $ orbits originating within the attractor remain there
and exhibit fully-developed aging properties, whereas if $t\rightarrow
\infty $ is taken before $\sigma \rightarrow 0$ a chaotic orbit with
exponential sensitivity to initial conditions would be observed.

\section*{Acknowledgments}
We thank an anonymous Referee for essential comments that allowed us to improve
considerably this work. 
FB kindly acknowledges hospitality at UNAM where part of this work has been
done. Work partially supported by DGAPA-UNAM and CONACyT (Mexican Agencies).

\end{document}